\documentclass[%
 aip,
 jmp,%
 amsmath,amssymb,
 reprint,%
]{revtex4-1}

\usepackage{graphicx}
\usepackage{dcolumn}
\usepackage{bm}

\begin{document}

\title{Palatini approach to modified $\mathbf{f(R)}$ gravity and its bi-metric structure}

\author{Janilo Santos}
\email{janilo@dfte.ufrn.br}
\author{Crislane de Souza Santos}
\email[]{crislane@dfte.ufrn.br}
\affiliation{Departamento de F\'{\i}sica, Universidade Federal do Rio G. do Norte,  59072-970 Natal - RN, Brazil}
\date{\today}%

\begin{abstract}
$f(R)$ gravity theories in the Palatini formalism has been recently used as an alternative
way to explain the observed late-time cosmic acceleration with no need of invoking either
dark energy or extra spatial dimension. However, its applications have shown that some subtleties
of these theories need a more profound examination. Here we are interested in the conformal aspects
of the Palatini approach in extended theories of gravity. As is well known, extremization of the
gravitational action {\it a la} Palatini, naturally ``selects'' a new metric $h$ related to the
metric $g$ of the subjacent manifold by a conformal transformation. The related conformal function is
given by the derivative of $f(R)$. In this work we examine the conformal symmetries of the
flat ($k=0$) FLRW spacetime and find that its {\it Conformal Killing Vectors} are directly linked to
the new metric $h$ and also that each vector yields a different conformal function.
\end{abstract}
\maketitle

\section{Introduction}

Understanding the physical mechanism behind the late-time cosmic acceleration is today one of the key problems at the interface between fundamental physics and cosmology. Much efforts in the last ten years have shown, at least in principle, that this phenomenon
could be explained by modifying Einstein's General Relativity (GR) in the far infrared regime. Although there is a diversity of approaches in this field (see, e.g., Refs.~\onlinecite{Modified_Gravity}), the simplest possible theory results by adding terms proportional to powers of the Ricci scalar $R$ to the Einstein-Hilbert Lagrangian. Such theory, known as modified $f(R)$ gravity (see Refs.~\onlinecite{Sotiriou} for recent reviews), is of great interest nowadays.
The cosmological interest in $f(R)$ gravity comes from the fact that it can exhibit naturally an accelerating expansion phase without introducing dark energy, as happens for instance in the standard $\Lambda$CDM cosmology. Much studies have been developed so far, mainly from the theoretical viewpoint, in order to clarify subtleties of this theory such as nonlocal causal structure~\cite{causal_structure}, its behavior under the energy conditions~\cite{energy_conditions} and Noether symmetries~\cite{Noether_symmetries}.

An important aspect that is worth emphasizing concerns the two different variational approaches that may be followed when one works with $f(R)$ gravity, namely, the metric  and the Palatini formalisms. In the metric formalism the connections are assumed to be the Christoffel symbols and the variation of the action is taken with respect to the metric, whereas in the Palatini variational approach the metric and the affine connections are treated as independent fields and the variation is taken with respect to both (see Ref.~\onlinecite{Olmo} for a quite good review on Palatini formulation). Because in the Palatini approach the connections depend on the particular $f(R)$, while in metric formalism the connections are defined {\it a priori} as the Christoffel symbols, the same $f(R)$ lead to different spacetime structures. In fact, these approaches are only equivalents in the context of GR, i.e., in the case of linear Hilbert action; for a general $f(R)$ term in the action, they provide completely different theories, with very distinct equations of motion.
Although being mathematically  more simple and successful in passing cosmological tests, we do not yet have a clear comprehension of the properties of the Palatini formulation of $f(R)$ gravity.

Thanks to Levi-Civita~\cite{levicivita} we know that every manifold may be endowed at least with two distinct structures: a Riemannian metric structure and an affine structure defined by the connection. The metric determines distances and local causality while the connection determines the parallel transport of vectors. In principle these structures are independent fields.
In metric $f(R)$ gravity, as well as in GR, a single geometric object, namely the metric $g$, determines the causal structure, measurements and the free-fall of test particles. However, in Palatini $f(R)$ gravity, as we will show in the next section, extremization of the gravitational action naturally selects $(i)$ a connection $\Gamma$, which is dependent of the particular $f(R)$ and $(ii)$ a second metric $h$ related to the metric $g$ of the manifold by $h_{\mu\nu}=f'g_{\mu\nu}$ ($f'$ is the derivative of $f(R)$). We are thus faced, in the Palatini approach, with the following questions:
\begin{itemize}
\item  What is the role of the connection $\Gamma$, as well as the second metric $h$, in Palatini $f(R)$ theory?
\item In a given manifold ($M,g$), is there some evidence of the ``apparent''  metric $h$?
\item If so, how many $h=f'g$ exist for a given $g$? In other words, a given function $f(R)$ is as good as any other?
\end{itemize}
The first question was raised and discussed in Refs.~\onlinecite{Capozziello}, where it was shown that $\Gamma$ is determined by the apparent metric $h$ and
the authors claim that the gravitational field should be represented by $\Gamma$, in the sense that free-falls follow the $\Gamma$-geodesics.
In what follows we try to answer the last two questions following geometrical reasonings concerned with symmetries present in the pair ($M,g$). Readers interested in observational (cosmological) aspects of Palatini $f(R)$ gravity see Refs.~\onlinecite{cosmological_aspects} and references therein.


\section{Palatini Approach to Modified $\mathbf{f(R)}$ Gravity}


The action that defines an $f(R)$ gravity is given by
\begin{equation}
\label{actionJF}
S = \int d^4x\sqrt{-g}\left[ \frac{f(R)}{2\kappa^2} + \mathcal{L}_m \right]\,,
\end{equation}
where $\kappa^2=8\pi G$, $g$ is the determinant of the metric tensor, $\mathcal{L}_m$ is the Lagrangian density for the matter
fields (here assumed as functionally independent of the connections $\Gamma$) and
\begin{equation}
\label{Ricc}
R = g^{\mu\nu}\left(\partial_{\alpha}\Gamma^{\alpha}_{\mu\nu} - \partial_{\nu}\Gamma^{\alpha}_{\mu\alpha} + \Gamma^{\alpha}_{\alpha\sigma}\Gamma^{\sigma}_{\mu\nu} - \Gamma^{\alpha}_{\nu\sigma}\Gamma^{\sigma}_{\mu\alpha} \right).
\end{equation}
In Einstein's General Relativity, as well as in metric $f(R)$ theories, the connections are given {\it a priori} as the Christoffel symbols of the metric $g$:
\begin{equation}
\label{Ein-approach}
\left\{^{\alpha}_{\mu\nu}\right\}_g =\frac{1}{2}g^{\alpha\sigma}\left( \partial_{\mu}g_{\sigma\nu} + \partial_{\nu}g_{\sigma\mu} - \partial_{\sigma}g_{\mu\nu} \right).
\end{equation}
In the Palatini variational approach metric and connections are treated as independent fields and the variation is taken with respect to both, giving us:
\begin{equation}  \label{eq_motion}
f' R_{(\mu\nu)}- \frac{1}{2}\,f\, g_{\mu\nu}=8\pi G\, T_{\mu\nu}\,,
\end{equation}
\begin{equation} \label{eq_connection}
\widetilde{\nabla}_\alpha\left( f'\sqrt{-g}\,g^{\mu\nu}\right)=0\,,
\end{equation}
where $f'=df/dR$, $\widetilde{\nabla}_\alpha$ denotes the covariant derivative
associated with $\Gamma$ and $T_{\mu\nu} = -(2\,/\!\sqrt{-g})\,\,\delta (\sqrt{-g}\mathcal{L}_m )
/ \, \delta g^{\mu\nu}$ is the matter energy-momentum tensor.   Eqs. (\ref{eq_motion}) are the modified Einstein's equations of motion, while solution of Eqs. (\ref{eq_connection}) give us the connections:
\begin{equation} \label{connections}
\Gamma^{\alpha}_{\mu\nu} = \frac{1}{2}h^{\alpha\sigma}\left( \partial_{\mu}h_{\sigma\nu} + \partial_{\nu}h_{\mu\sigma} - \partial_{\sigma}h_{\mu\nu} \right)\,,
\end{equation}
where $h_{\mu\nu}=f'g_{\mu\nu}$ is a new {\it conformal metric}. Let us remark that:
\begin{itemize}
\item The dynamics of $\Gamma$, expressed by Eq.~(\ref{eq_connection}), identifies a new metric $h_{\mu\nu}$ in the manifold,
and this is the connection which determines the tensor curvature of spacetime.
\item The conformal metric $h$ preserves the causal structure of the manifold ($M,g$).
\item  If $f(R)=R$ the field equations (\ref{eq_connection}) $\Rightarrow$
            $\Gamma^{\alpha}_{\mu\nu} = \left\{^{\alpha}_{\mu\nu}\right\}_g$, so the Levi-Civita connection of $g$ (Eq. (\ref{Ein-approach})) is no longer an assumption {\it a priori}, it is the outcome of field equations!
\end{itemize}
Next we are going to explore the conformal character of the new metric $h$.


\section{Conformal Transformations $\&$ Lie Derivatives}


As stated above, let ($M,g$) be a spacetime with a smooth Lorentzian metric $g$. A symmetry of the spacetime $M$ is a smooth local diffeomorphism $\phi_t$ between open submanifolds of $M$ which preserves some geometrical feature of $M$. The symmetries are usually achieved by assuming the existence of smooth vector fields on $M$ associated with the local diffeomorphism. These vectors are called ``symmetry vector fields'' (see Ref.~\onlinecite{hall} for a detailed account of algebraic structures in general relativity). Much known in GR are the Killing vector fields, whose associated local diffeomorphism preserve the metric tensor. Another important symmetry is related to the so-called ``Conformal Vector Fields''.  We say that ${\bf X}$ is a {\it Conformal Vector Field} (also called {\it Conformal Killing Vector}) if the  associated diffeomorphisms $\phi_t$ preserves the metric up to a conformal factor: $\phi_t^{\ast}g = \Omega(x)g$ for some positive function $\Omega$.
This can be cast in terms of the Lie derivative of the metric tensor as~\cite{Schouten,Yano}
\begin{equation}  \label{Conformal_symm}
\textrm{\pounds}_{{\bf X}_A}\,g_{\mu\nu} = 2\phi_A g_{\mu\nu}
\end{equation}
where $\phi_A$ is called the conformal function of ${\bf X}_A$ and
\begin{equation}
\textrm{\pounds}_{{\bf X}}\,g_{\mu\nu} = X^{\alpha}\partial_{\alpha}g_{\mu\nu}+g_{\alpha\nu}\partial_{\mu}X^{\alpha} +  g_{\mu\alpha}\partial_{\nu}X^{\alpha}\,.
\end{equation}

Here we are interested in the {\it Conformal Vector Fields} ${\bf X}_A$, and respective conformal functions, possibly admitted by the flat ($k=0$) Friedmann-Lema\^{i}tre-Robertson-Walker (FLRW) spacetime such that
\begin{equation} \label{Conformal_Factor}
\textrm{\pounds}_{{\bf X}_A}\,g_{\mu\nu} = 2\,f'_A g_{\mu\nu}
\end{equation}
where $f'_A\neq 0$ are derivatives of $f(R)$ Palatini theories of gravity.

\subsection{Conformal Vector Fields in ($k=0$) FLRW}

We write the flat FLRW spacetime metric as
\begin{equation} \label{Friedmann_metric}
ds^2 = a^2(\eta)\left( - d\eta^2 + dx^2 + dy^2 + dz^2 \right),
\end{equation}
where $\eta$ is the conformal time; $d\eta = dt/a(t)$. It is well known~\cite{choquet,Maartens} that for this metric there are 9 {\it Conformal Vectors}:
\begin{equation}
\begin{array}{llr}
{\bf X}_0 = \partial_{\eta} &  &  \\
{\bf X}_i = x_i\,\partial_{\eta} + \eta\,\partial_i & &  \\
{\bf X}_4 = x^{\alpha}\partial_{\alpha} & & \\
{\bf K}_{0}= -2\eta\,{\bf X}_4 -(x_{\alpha}x^{\alpha}){\bf X}_0 &  & \\
{\bf K}_{i}= 2x_i\,{\bf X}_{4} & &
\end{array}
\end{equation}
where $x^{\alpha}=(\eta,x,y,z)$,  $x_{\alpha}=(-\eta,x,y,z)$ and $i=1,2,3$. Using the above results in
Eqs. (\ref{Conformal_Factor})-(\ref{Friedmann_metric}) we determine the ``conformal functions'' $f'_A(R)$ associated with each {\it Conformal Vector Field} ${\bf X}_A$ (${\small A} =1\ldots 9$). The results are shown in Table~\ref{conformal_factor} where we see that, in principle, there are 9 different $f(R)$ theories associated with the conformal symmetries of the flat FLRW spacetime. Although they were obtained in terms of the derivative of $f(R)$, and in spite of integration being not obvious, we recall that many discussions and constraints on $f(R)$ theories are made in terms of the derivatives. Indeed, these are preliminary results which we pretend to develop more deeply in forthcoming studies.
\begin{table}
\begin{tabular}{|l|l|r|} \hline\hline
{\it Conformal Vector} & $f'=df/dR$ ({\it associated $f(R)$ theory})  \\  \hline
${\bf X}_0$ & ${\bf d\ln{a}/d\eta}$ \\  \hline
${\bf X}_i$ & ${\bf x_i\,d\ln{a}/d\eta}$ \\  \hline
${\bf X}_4$ & ${\bf 1 + \eta\,d\ln{a}/d\eta}$ \\  \hline
${\bf K}_0$ & ${\bf - 2\eta -(\eta^2 + x^2 + y^2 + z^2)\,d\ln{a}/d\eta}$ \\  \hline
${\bf K}_i$ & ${\bf 2\,x_i\,(1 + \eta\,d\ln{a}/d\eta)}$ \\  \hline\hline
\end{tabular} 
\caption{\label{conformal_factor} The right column shows $f(R)$ Palatini theories (indeed $f'(R)$) associated with each Conformal Killing Field (left column) of the FLRW metric.}
\end{table}


\section{Conclusions}


In this talk we have shown that:
\begin{itemize}
\item In the Palatini approach to $f(R)$ gravity, beside the metric $g$, another metric $h$ is involved, which gives the connection $\Gamma$. These two metrics are related by a conformal
transformation such that $h_{\mu\nu}=f'g_{\mu\nu}$.
\item  The new metric $h_{\mu\nu}$, as well as the connection $\Gamma$, is directly linked to {\it Conformal Killing} symmetries preexistent in the manifold ($M,g$).
\item We relate the {\it Conformal Killing Vectors} in FLRW flat spacetime to $f(R)$ theories of gravity in the Palatini approach (see Table~\ref{conformal_factor}).
\end{itemize}
We hope this result will enlighten the role of conformal transformations, present in the Palatini formulation {\it ab initio}, and the related bi-metric structure of these theories.

\begin{acknowledgments}

The authors wish to acknowledge financial support from Brazilian Agencies: J. Santos acknowledge support from CNPq and C.S. Santos the support of CAPES/REUNI.

\end{acknowledgments}

\end{document}